\documentclass[12pt]{article}
\usepackage{epsfig}
\begin{document}

\title{Nonstationary time series analysis of\\
    heart rate variability}
\author{P.B. Siegel$^1$, J. Sperber$^2$, W. Kindermann$^2$,
   and A. Urhausen$^2$}
\maketitle
\noindent $^1$Physics Department, California State 
Polytechnic University Pomona, Pomona, CA 91768\\
$^2$Institute of Sport and Preventative Medicine, University of 
Saarland, 66041 Saarbruecken, Germany

\begin{abstract}
An analysis of the RR-interval time series, $t_i$, is presented for
the case in which the average time, $\bar{t}$, changes slowly.  
In particular, $\bar{t}$ and a short-time scale variability
parameter, $V$, are simultaneously measured while $\bar{t}$ decreases
for subjects in the reclined position.
The initial decrease in $\bar{t}$ is usually linear with
$V$ yielding parameters that can be related to
physiological quantities.
\end{abstract}

\centerline{PACS: 87.19.Hh, 87.80.Tq, 87.80.Vt, 05.45.Tp}

\section{Introduction}

   Heart rate variability is used to examine the
autonomous nervous system's control of heart rate, with the goal
of assessing health and fitness in humans\cite{eckberg}.  The main
measurement is the time between successive heartbeats, which
is referred to as the RR-interval.  Measurements are taken
over time periods that can be as short as minutes or as
long as days, and the resulting data are a series of times, $t_i$,
measured to an accuracy of milliseconds.  Data analysis includes
methods from spectral analysis\cite{prok, eckberg2, malik, pagani},
statistical physics
\cite{costa, yulm}, and nonlinear dynamics\cite{peng, thurner, wessel, ching}.  
These methods are applied to situations for which the time
series is stationary, that is,
the average value of the $t_i$ does not change significantly in time.
These approaches have been useful
in determining certain aspects of heart rate control.  However, there may
be ambiguities in relating the analysis parameters to physiological quantities.  

The purpose of this article is to consider the feasibility of using a
controlled nonstationary time series
of the RR-interval times to assess heart rate control.  If
the average value of the $t_i$, $\bar{t}$, changes sufficiently
slowly, then it
is possible to also measure a short-time scale variability
parameter, $V$.  
If $\bar{t}$ and $V$ depend upon physiological
quantities in different ways, then a plot in the $\bar{t} - V$
plane for a controlled process may yield information about the
influences causing the change.  Although we use a specific
measurement for $V$, the respiratory sinus arrhythmia amplitude,
the approach can be applied for any measure of $V$.

\section{Description of Measurements and Data Analysis}

The average heart period ($\bar{t}$) and the variability ($V$) of the
RR-interval times depend primarily on the
autonomous nervous system, which has two main
influences: sympathetic ($m$) and parasympathetic ($n$) nerve activity.  
One would like to learn about $m$ and $n$ from the non-invasive
measurements of $\bar{t}$ and $V$.  A model used
by physiologists to describe the relationship of the average heart
rate $B$ to the sympathetic and parasympathetic influences is given by
the equation \cite{katona, fouad, kenney} $B = B_0 m n$
where $B_0$ is the intrinsic heart rate, $m$ is the sympathetic
factor and $n$ the parasympathetic factor.  The sympathetic
factor $m$ is greater than one, increasing the heart rate, and
the parasympathetic factor $n$ is less than one, decreasing the
heart rate.  The factors $m$ and $n$ which multiply the intrinsic heart rate 
depend, among other influences, upon a person's activity level, posture, recent 
diet, degree of fitness, age, and health.  We label all 
these parameters, which affect the sympathetic (m) and parasympathetic (n)
factors, by the symbol $\beta$: m($\beta$) and n($\beta$).  One can 
also write $n=1/(1+P(\beta))$ \cite{katona}, where the parasympathetic
control parameter $P$ is unitless and varies from zero to its maximum 
value $P_{max}$.  In Ref. \cite{katona} $P$ has units of milli-seconds.
In terms of the average heart period, the equation can 
be written as $\bar{t} = T_0(1+P(\beta))/m(\beta)$, where 
$T_0 = 1/B_0$ is the intrinsic heart period.

While in general variability is influenced by both sympathetic and
parasympathetic effects, short-time scale variability depends primarily
on parasympathetic activity.  Short-time scales are
variations with frequencies greater than $0.15$ Hz.  Since
average heart rates are around one beat per second, this corresponds
to variations in heart rate taking place in less than roughly 6 heartbeats.
There is good evidence that high frequency ($>0.15$Hz) spectral power 
depends only on $P$, and is not influenced by sympathetic nerve
activity.  Thus, if $V$ is a measure of short-time
scale variation, it can be expressed as a function of parameters
other than $m$, namely $V(P, \beta)$.
Expressing $P$ in terms of $V$ and the "$\beta$" factors, 
the relationship between the
average heart period and its variability can be written as
$\bar{t} = T_0 (1+P(V, \beta))/m(\beta)$.  Measurements
of $\bar{t}$ and $V$ during slow changes of these values can 
yield information about changes in $m$, $T_0$, and the 
function $P(V, \beta)$.

For this study, we chose as the variability parameter $V$ the
respiratory sinus arrhythmia (RSA) amplitude,
which is the amplitude of the fluctuations of the RR-interval times 
due to breathing, for a breathing rate of 12 breaths/min.
We have chosen this parameter for $V$ since it has been shown in
unconscious dogs\cite{katona} and humans\cite{fouad} that
($\bar{t} - T_0/m$) is proportional to the RSA amplitude.  
The experiments carried out in Refs. \cite{katona} and \cite{fouad} 
used pharmacological methods (parasympathetic blockade) to decrease $\bar{t}$ 
while keeping $m$ constant.  If this linear relationship
holds in general, then $V(P, \beta)$ takes on the simple
form $V = k(\beta)P$, where k has the same units as $V$. 
Letting $V$ represent the RSA amplitude for a 12 breath/min breathing 
rate, we have

\begin{equation}
  \bar{t} = {T_0 \over m} (1 + {{V} \over {k}} ) 
\end{equation}

\noindent where $m$ and $k$ might depend on posture, tidal volume, 
environmental factors, diet before measurement, etc.  The hope is that 
$\bar{t}$ and $V$ can be varied keeping $k$ constant.

We measured both the average heart period in msec, and the amplitude
of the RSA in
msec for various subjects while they underwent slow warm up in the lying 
position.  Since the amounts of sympathetic and parasympathetic influences
are posture dependent, it is important to maintain the same posture
while the heart rate changes.  The lying position was chosen because
the parasympathetic activity is maximized in this position.
The subject slowly increased his/her pulse
by pedaling a stationary bicycle in the lying position while breathing
at 12 breaths/min the whole time.  He/she started by lying still
for 2 minutes, and then began pedaling slowly with a load of 25 
watts for the duration of one minute.  The subject then pedaled faster until
the heart rate increased to around 20 bpm above its resting value.  The
load was increased by increments of 25 watts each two minutes
thereafter, until a pulse of around 110 beats/min was reached.  

Breathing was controlled by having the subjects breathe in concert
with increasing and decreasing tones, which they listened to from a 
CD player.  RR interval times were recorded using a heart rate 
monitor from Polar, Model S810.  The average heart period 
and RSA amplitude were computed from a 
sequence of 100 consecutive RR-interval times, with the start of the
interval shifted 50 beats after each calculation.  For example, the first
sequence is beat numbers 1 to 100, the second from 50 to 150, etc.
Since breathing was
controlled at 12 breaths/min, the peak in the Fourier spectrum
was distinct at 0.2 Hz.  
From each sequence of 100 times, the RSA amplitude was set equal to this
peak Fourier amplitude at or near the breathing frequency of 
12 cycles/min (i.e. 0.2 Hz).

\section{Discussion}

We carried out the test for 20 subjects, and plot some
typical cases in figures 1 to 3.  In each figure, the initial part of the warm
up is linear, labelled $1 \rightarrow 2$, in
which both the heart period and the variability decrease.
This is followed by a decrease in heart period with little change
in variability, labelled as $2 \rightarrow 3$ in the figures.  About
$75\%$ of the subjects we measured had similar plots.

The data can be interpreted using the model of Eq. 1.  During the initial linear
stage of the warm-up, $1 \rightarrow 2$, parasympathetic influences
are reduced since the RSA amplitude decreases.  If the RSA amplitude
is proportional to $P$, the data are consistent with  
$T_0/m$ and $k$ being constant during this first stage.  After the 
initial parasympathetic decrease,
the plot moves to the left, $2 \rightarrow 3$.  Since the variability does 
not change during this stage, the data indicate that $m$ is increasing.
For the initial linear stage, three parameters can
be determined: $T_0/m$, $k$, and $P_{max}$.  The parameter $P_{max}$
is unitless,
and is the fraction that the parasympathetic activity increases the 
heart period from $T_0$ to the maximum value of $\bar{t}$:
$P_{max} = (t_{max} - T_0/m)/(T_0/m)$, where
$t_{max}$ is the maximun value of the average heart period.
These parameters can always be determined for a linear stage of the graph.
If the model of Eq. 1 is valid and $k$ remains constant, then 
the parameters have physiological significance.
In the figures we state $P_{max}$, $k$, and $T_0/m$ for the
slow warm-up phase.

Some subjects show interesting results, which we graph in figures 2 and 3.  In
figure 2, the plot shows four stages which can be interpreted using Eq. 1.  
Initially ($1 \rightarrow 2$) the RSA amplitude decreases significantly
suggesting $P$ decreases with $m$ constant.  Then ($2 \rightarrow 2a$) 
the heart period decreases but the RSA amplitude remains constant, meaning
that $m$ increases with $P$ constant.  The next change from $2a \rightarrow 2b$ 
is similar to the change from $1 \rightarrow 2$, and the final stage from
$2b \rightarrow 3$ has $m$ increasing with $P$ constant as from
$2 \rightarrow 2a$.  For this subject it appears that the heart rate
increases via alternating parasympathetic and sympathetic influences.

In figure 3 we compare two different subjects.  Both subjects have the
same resting heart rate of 50 bpm.  One subject (circles) has a resting RSA amplitude
of 70 msec, while the other (triangles) only 25 msec.  One might be lead to believe 
that the first subject has a stronger parasympathetic activity, since the
RSA amplitude is larger.  However, during warm-up the RSA amplitude for
the first subject (circles) was reduced to its minimum value of 10 msec 
for only a 350 msec decrease in
heart period.  For the other subject (triangles), the heart period
decreased by 600 msec
before the RSA amplitude was reduced to its minimum value.  Thus the 
second subject has a larger change in heart period due to the reduction
in parasympathetic activity, and consequently a larger value of $P_{max}$.
The plots in the $\bar{t}-V$ plane are quite different for these
subjects even though they have the same resting heart rate.  

We also used the absolute value of the difference between successive
RR-intervals without controled breathing as a variability parameter.
If the subject is breathing
at a consistent rate throughout the exercise, the results are
similar to those obtained using the RSA amplitude
with a different value of $k$.  Other possible
candidates for a variability parameter are the standard deviation of the 
RR-interval times, the square-root of the 
high-frequency power, or the RSA from Ref. \cite{hatfield}.
Heart rate ($1/\bar{t}$) versus
variability are plotted in Refs. \cite{hatfield} and \cite{grossman}.
However, in neither case are the data modeled using the parameters of
Eq. 1.  In figure 4 of Ref. \cite{grossman}, the loci of points are similar
to the figures presented here: a large decrease in RSA for a small increase
in heart rate (parasympathetic reduction), and a small decrease in
RSA for a large increase in heart rate (sympathetic increase).  Perhaps
if the time ordering of the points and posture are taken into account, 
$T_0/m$, $P_{max}$, and $k$ could be measured for different types of daily
activity.

In conclusion, we propose that a simultaneous measurement of the
average heart period $\bar{t}$ and a parameter related to short time-scale variability 
$V$ can yield insight into the autonomous nervous system.  For slow
changes in the heart rate, a graph in the $\bar{t}-V$ plane can assist
in understanding the influences causing the change.  For situations
in which the change in heart rate is caused only by the parasympathetic
nerve activity, one can extract three parameters, $T_0/m$, $P_{max}$, 
and $k$.  These parameters could have a simple relationship to 
physiological quantities.

\newpage

\begin{center}
{\bf Figure Captions}
\end{center}

\bigskip

\noindent Figure 1.  Graph of RSA (respiratory sinus arrhythmia)
versus heart period as a subject's heart period slowly deceases while
in the lying position.  From the straight-line section $1 \rightarrow 2$ 
the parameters $T_0/m$, $P_{max}$, and $k$ are determined using equation 1.

\bigskip

\noindent Figure 2.  Same graph as in Figure 1, but for a different 
subject.  The decrease in heart period can be interpreted as alternating
between the influences of parasympathetic decrease ($1 \rightarrow 2$) 
and ($2a \rightarrow 2b$) and sympathetic increase ($2 \rightarrow 2a$) 
and ($2b \rightarrow 3$).
  
\bigskip

\noindent Figure 3.  Same graph as in Figure 1, comparing two different 
subjects.  Both subjects have the same resting rate.  Eventhough one
subject (circles) has a higher RSA, the parasympathetic control 
parameter $P_{max}$ is less.  

\newpage

\begin{figure}
\epsfxsize=16cm
\epsfbox{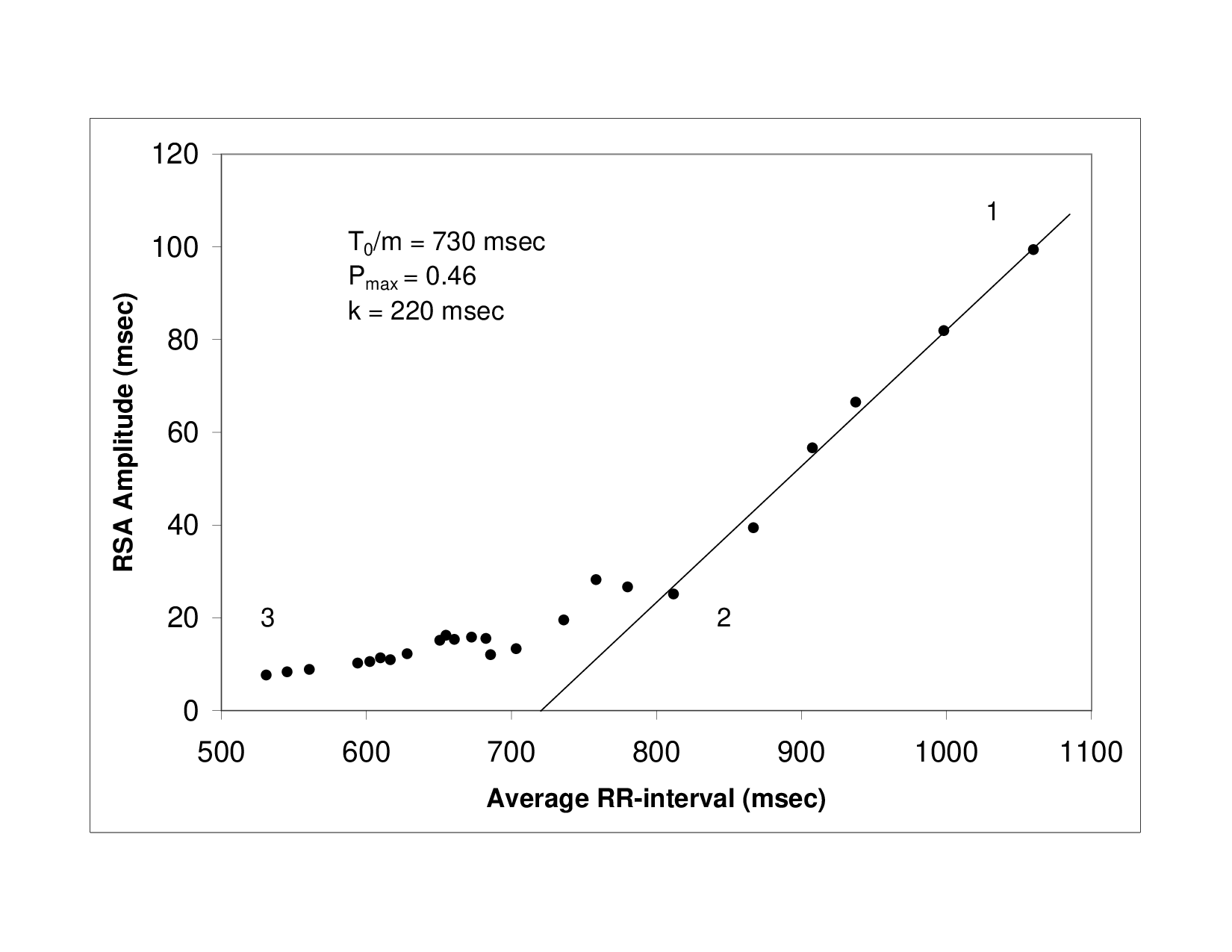}
\end{figure}

\noindent Figure 1.  Graph of RSA (respiratory sinus arrhythmia)
versus heart period as a subject's heart period slowly decreases while
in the lying position.  From the straight-line section $1 \rightarrow 2$ 
the parameters $T_0/m$, $P_{max}$, and $k$ are determined using equation 1.

\newpage

\begin{figure}
\epsfxsize=16cm
\epsfbox{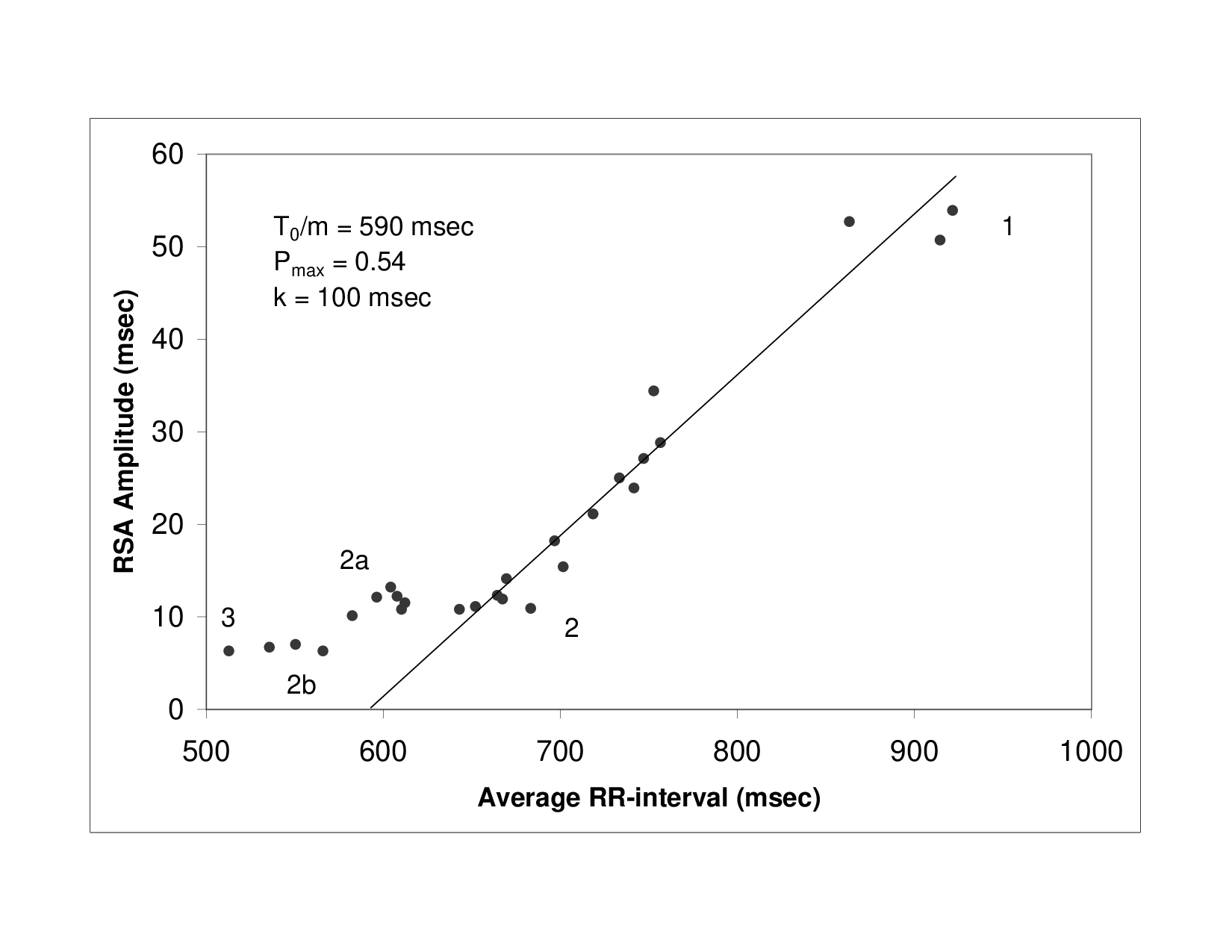}
\end{figure}

\noindent Figure 2.  Same graph as in Figure 1, but for a different 
subject.  The decrease in heart period can be interpreted as alternating
between the influences of parasympathetic decrease ($1 \rightarrow 2$) 
and ($2a \rightarrow 2b$) and sympathetic increase ($2 \rightarrow 2a$) 
and ($2b \rightarrow 3$).

\newpage

\begin{figure}
\epsfxsize=16cm
\epsfbox{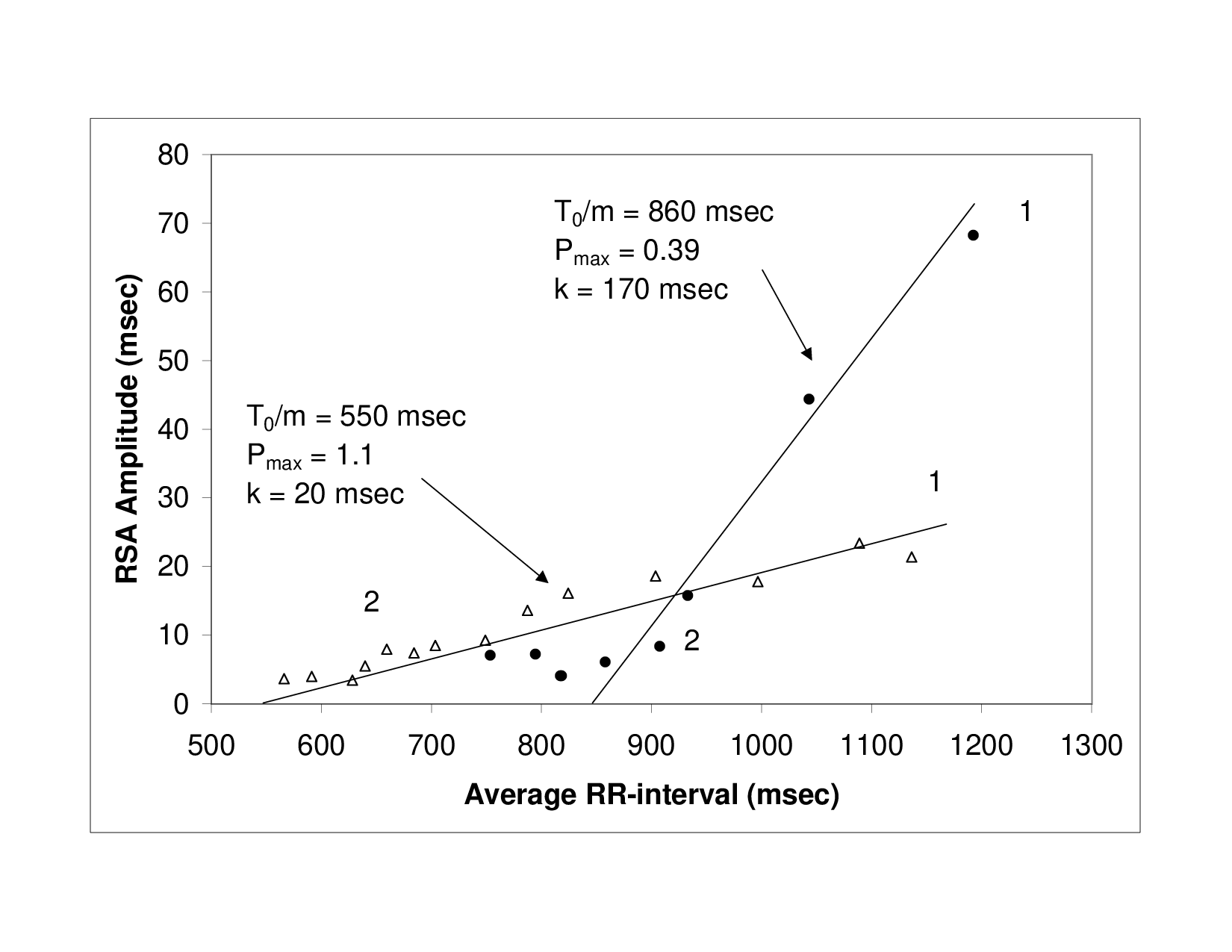}
\end{figure}

\noindent Figure 3.  Same graph as in Figure 1, comparing two different 
subjects.  Both subjects have the same resting rate.  Eventhough one
subject (circles) has a higher RSA, the parasympathetic control
parameter $P_{max}$ is less.  


\begin{thebibliography}{hghg}

\bibitem{eckberg} D. Eckberg, Ann. Med. {\bf 32}, 341-349 (2000).

\bibitem{prok} M.D. Prokhrov, V.I. Ponomarenko, V.I. Gridnev, M.B. Bodrov,
and A.B. Bespyatov, Phys. Rev. E {\bf 68}, 041913 (2003).

\bibitem{eckberg2} D. Eckberg, Circulation {\bf 96}, 3224-3232 (1997).

\bibitem{malik} M. Malik, Circulation {\bf 93}, 1043-1065 (1996).

\bibitem{pagani} M. Pagani et. al., Circulation Research {\bf 59}, 
178-193 (1986).

\bibitem{costa} M. Costa, A.L. Goldberger, and C.-K. Peng, Phys. 
Rev. Lett. {\bf 89} 068102 (2002).

\bibitem{yulm} R. Yulmetyev, P. Hanggi, and F. Gafarov, Phys. Rev.
E {\bf 65}, 022901 (2002).

\bibitem{peng} C.-K. Peng, J. Mietus, J>M> Hausdorff, S. Havlin, H.E. 
Stanley, and A.L. Goldberger, Phys. Rev. Lett. {\bf 70}, 1343-1346 (1993).

\bibitem{thurner} S. Thurner, M.C. Feurstein, and M.C. Teich, Phys. 
Rev. Lett. {\bf 80}, 1544-1547 (1998).

\bibitem{wessel} M. Wessel, A. Schirdewan, and J. Kurths, Phys. Rev.
Lett. {\bf 91}, 119801 (2003).

\bibitem{ching} E. Ching, D.C. Lin, and C. Zhang, Phys. Rev. E 
{\bf 69}, 051919 (2004).

\bibitem{katona} P. Katona and F. Jih, J. Appl. Physiol. {\bf 39}, 
801-805 (1975).

\bibitem{fouad} F. Fouad, R. Tarazi, C. Ferrario, C Fighaly 
and C. Alicandri, Am. J. Physiol. {\bf 246}: H838-H842 (1984).

\bibitem{kenney} W. Kenney, Med. Sci. Sports. Exer. {\bf 17},
451-455 (1985).

\bibitem{hatfield} B. Hatfield, T. Spalding, 
D. Santa Maria, S. Porges, J. Potts, E. Byrne, E. Brody, and 
A. Mahon, Med. Sci. Sports Exerc. {\bf 206}, 14 (1998).

\bibitem{grossman} P. Grossman, F. Wilhelm, and M. Spoerle, Am. J. 
Physiol.Heart Circ. Physiol. PMID: 14751862 (2004).

\end{thebibliography}
\end{document}